\documentclass[sn-basic]{sn-jnl}
\usepackage{multirow}
\usepackage{amsmath,amssymb,amsfonts}%
\usepackage{amsthm}%
\usepackage{mathrsfs}%
\usepackage[title]{appendix}%
\usepackage{xcolor}%
\usepackage{textcomp}%
\usepackage{manyfoot}%
\usepackage{booktabs}%
\usepackage{algorithm}%
\usepackage{algorithmicx}%
\usepackage{algpseudocode}%
\usepackage{listings}%
\usepackage{indentfirst}
\usepackage{amsmath}
\usepackage{graphicx,psfrag,epsf}
\usepackage{graphicx}
\usepackage{enumerate}
\usepackage{natbib}
\usepackage{subfig}
\usepackage{mathrsfs}
\usepackage{algpseudocode}
\usepackage{caption}
\usepackage{subcaption}
\usepackage{amsthm,amssymb,amsmath,dsfont, mathtools}
\usepackage{algorithm, algorithmicx}
\usepackage{array}
\usepackage{tabularx}
\usepackage{nicematrix}
\usepackage{appendix}
\usepackage{ragged2e}

\newcommand {\vv}[1]{\mbox{vec}}

\newcommand{\bx}{\mathbf{x}}

\newcommand{\bW}{\mathbf{W}}

\newcommand{\be}{\mathbf{e}}

\newcommand{\bw}{\mathbf{w}}
\newcommand{\bz}{\mathbf{z}}

\newcommand{\by}{\boldsymbol{y}}

\newcommand{\calW}{\mathcal{W}}

\begin{document}

\title[Article Title]{A least distance estimator for a multivariate regression model using deep neural networks}

\author[1,2]{\fnm{Jungmin} \sur{Shin}}\email{c16267@kma.ac.kr}

\author[2]{\fnm{Seung Jun} \sur{Shin}}\email{sjshin@korea.ac.kr}

\author*[1]{\fnm{Sungwan} \sur{Bang}}\email{wan1365@gmail.com}

\affil[1]{\orgdiv{Department of Mathematics}, \orgname{Korea Military Academy}, \orgaddress{\street{574 Hwarang-Ro}, \state{Nowon-Gu}, \postcode{01805}, \city{Seoul}, \country{Korea}}}
\affil[2]{\orgdiv{Department of Statistics}, \orgname{Korea University}, \orgaddress{\street{145 Anam-Ro}, \state{Seongbuk-Gu}, \postcode{02841}, \city{Seoul}, \country{Korea}}}

\abstract{\hspace{.3cm} 
We propose a deep neural network (DNN) based least distance (LD) estimator (DNN-LD) for a multivariate regression problem, addressing the limitations of the conventional methods. Due to the flexibility of a DNN structure, both linear and nonlinear conditional mean functions can be easily modeled, and a multivariate regression model can be realized by simply adding extra nodes at the output layer. The proposed method is more efficient in capturing the dependency structure among responses than the least squares loss, and robust to outliers. In addition, we consider $L_1$-type penalization for variable selection, crucial in analyzing high-dimensional data. Namely, we propose what we call (A)GDNN-LD estimator that enjoys variable selection and model estimation simultaneously, by applying the (adaptive) group Lasso penalty to weight parameters in the DNN structure. For the computation, we propose a quadratic smoothing approximation method to facilitate optimizing the non-smooth objective function based on the least distance loss. The simulation studies and a real data analysis demonstrate the promising performance of the proposed method.}

\maketitle

\section{Introduction}\label{INTRO}
In conventional statistical learning, the response is often univariate. However, it is not uncommon to predict several variables simultaneously in practice. For instance, a researcher in a hospital wants to predict the physical conditions of the patient such as blood pressure, blood sugar rate, stage of a certain disease, etc. Of course, it is not natural to expect that those responses are independent. Therefore, we have taken into account their dependent structure to the model to improve the performance of the model. 
This naturally leads to the multivariate regression model as follows:
\begin{equation}\label{model}
\boldsymbol{y}=\boldsymbol{f(\bx)}+\boldsymbol{\epsilon},
\end{equation}
where $\by = (y_1, y_2, \cdots, y_q)^{\top} \in \mathbb{R}^q$, $\bx=(x_1, \cdots, x_p)^{\top} $ denote $q$- and $p$- variate response and predictor variable, respectively, with $\boldsymbol{\epsilon}=(\epsilon_1, \cdots, \epsilon_q)^{\top}$ being a random error vector centered at zero. Our goal is to estimate multivariate regression function $\boldsymbol{f}=(f_1, \cdots,f_q)^{\top}$ to reveal an association between $\by$ and $\bx$, where both are multivariate. A canonical choice in statistical communities is the multivariate linear regression model \citep[Among many others,][]{Breiman1997,Bai1990lde,Sohn2012} which assumes
\begin{equation}\label{linear_reg}
f_k\left(\bx\right) = \bw_k^T \bx, \quad k=1, \ldots, q,
\end{equation} 
where we slightly abuse the notation for the simplicity by letting $\bx=\left(1, \bx^T\right)^T$, and $\bw_k=\left(w_{k 0}, w_{k 1}, \cdots, w_{k p}\right)^{\top}$ denotes the coefficient vector associated with the $k$th response, $y_k$. Given $n$ independent copies of $(\bx, \by)$, the least square (LS) estimator solves 
\begin{equation}\label{linear_reg_est}
\widehat{\bw}_k^{\textrm{LS}}=\underset{w \in \mathbb{R}^{p+1}}{\arg \min } \sum_{i=1}^n\left(y_{i k}-\bw_k^T \bx_i\right)^2,  k=1, \cdots, q.
\end{equation}
Let, $\bW = \left[ \bw_1 \cdots \bw_k \cdots \bw_q\right]^{\top} = \left[ \bw_{(0)} \bw_{(1)} \cdots \bw_{(j)} \cdots \bw_{(p)}\right]$ denote the $q \times (p+1)$  coefficient matrix, where $\bw_{(0)}= (w_{10}, \cdots, w_{q0} )^{\top}$  and   $\bw_{(j)}= (w_{1j}, \cdots, w_{qj} )^{\top}$, $j=0, \cdots, p$ represent the coefficient vectors associated with the intercept and the $j$th predictor, respectively. Using the Euclidean norm denoted by $\left\| \boldsymbol{\cdot} \right\|$, the LS estimator \eqref{linear_reg_est} is simply rewritten as the minimizer of the sum of squared norm:
\begin{equation} \label{linear_reg_est2}
\widehat{\bW}^{\textrm{LS}}=\underset{\bW \in \mathbb{R}^{(p+1) \times q}}{\arg \min }  \sum_{i=1}^n\left\|\by_i-\bW \bx_i\right\|^2.
\end{equation} 

\begin{figure}[t]
\centering
    \subfloat[Neural networks with two hidden layers for a univariate regression model.]{{\includegraphics[width=5.1cm]{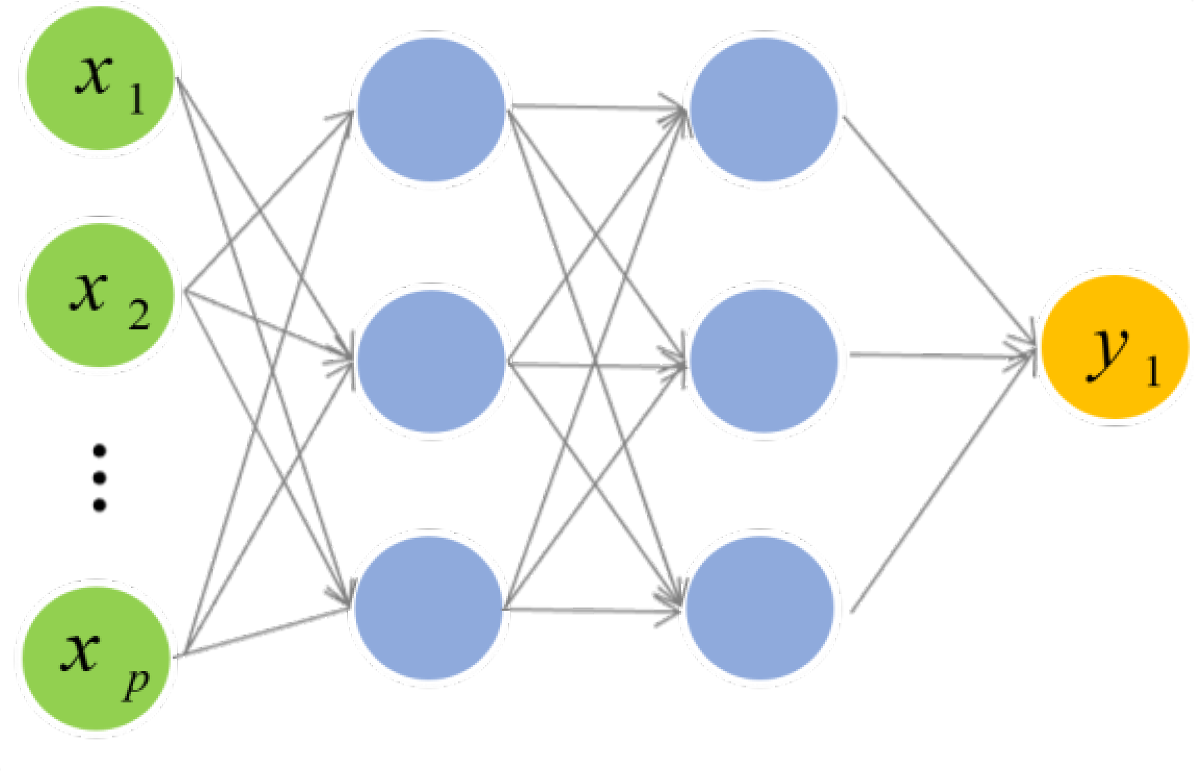} }}%
    \qquad
    \subfloat[Neural networks for a bivariate regression model by adding one more node in the output layer.]{{\includegraphics[width=5cm]{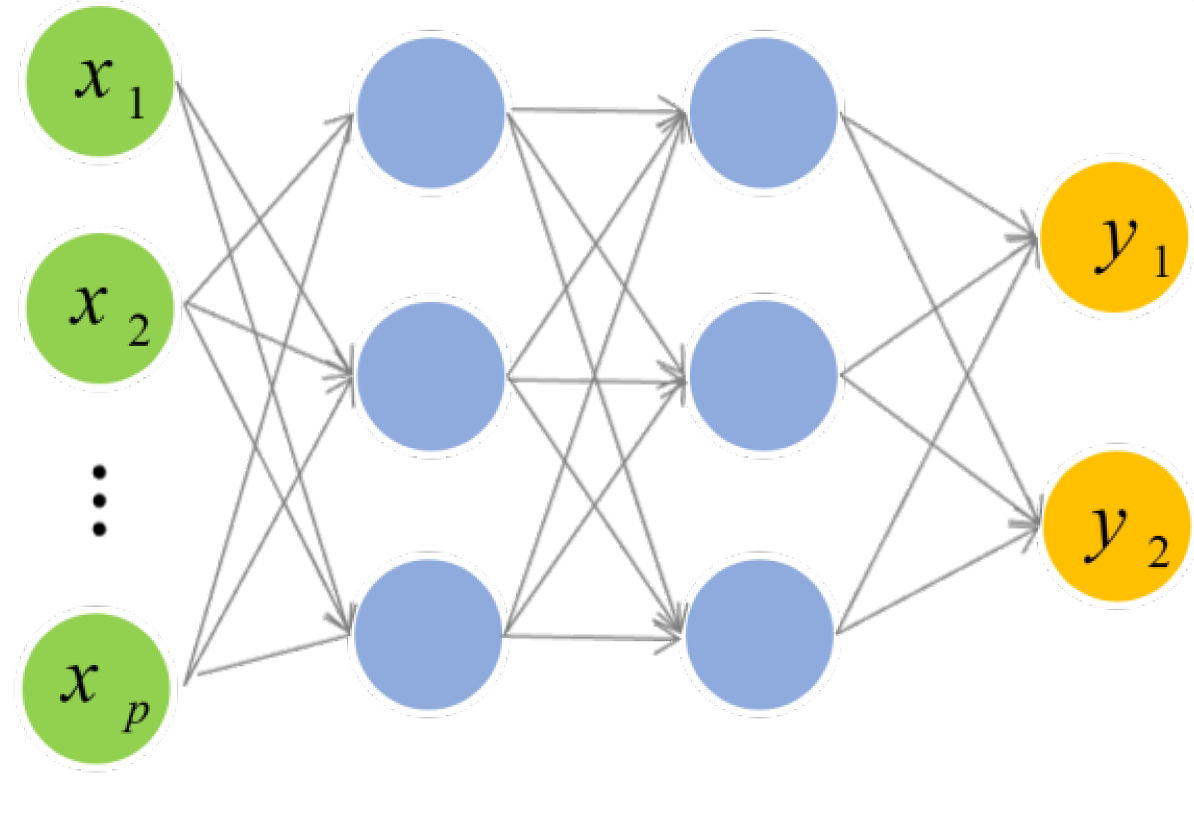} }}%
    \caption{Extensible property of a neural network for being multivariate regression model from a univariate one.}
    \label{fig:multivariate_reg}
\end{figure}

Despite of its popularity, the linearity model \eqref{linear_reg} is often too stringent in many applications, where the true relations are highly nonlinear. There are comprehensible ways in machine learning to extend the linear model to the nonlinear one, such as local regression, spline, and kernel trick based on reproducing kernel Hilbert space \citep{wahba1992multivariate}. In the meantime, the (deep) neural network (DNN) becomes a canonical choice in modeling complex data structure due to its promising performance and flexible structure \citep[among many others]{imaizumi2019deep,sze2017efficient,cao2020deep}. On top of that, DNN is very straightforward to extend from a univariate to multivariate response by simply adding more nodes at the output layer (Fig. \ref{fig:multivariate_reg}). 

In terms of estimation, the LS estimator defined in \eqref{linear_reg_est2} is popular in practice due to its computational and theoretical simplicity. However, it often suffers. For example, the LS estimator that essentially seeks the conditional mean of $\by$ given $\bx$, which is not the best choice under the presence of outliers. In addition, the multivariate responses are often highly correlated as the biomedical application given above, and the LS estimation could be inefficient to capture such dependency structure among responses since it solves the problem independently for each each coordinate. Even the LS estimator under the DNN framework can partly use correlation information between responses to optimize the weights between the layers. This will be illustrated more clearly in Section \ref{sec2}.

As a natural alternative to the least squares estimation, \cite{Bai1990lde} proposed least distance (LD) estimator that solves  
\begin{equation}\label{ld_reg}
\widehat{\bW}^{\textrm{LD}}=\underset{\bW \in \mathbb{R}^{(p+1)\times q}}{\arg \min } \frac{1}{n} \sum_{i=1}^n\left\|\boldsymbol{y}_i-\bW \mathbf{x}_i\right\|,
\end{equation}
under \eqref{linear_reg}.
That is, the LD estimator minimizes the Euclidean distance between the response vector and estimate from the model. Thus it is conceptually akin to the spatial median \citep{Haldane1948}, a popular alternative of multivariate mean, and thus \eqref{ld_reg} is robust under the presence of outliers. It is also more efficient to capture the dependency structure among responses compared to the LS estimator \citep{Bai1990lde,Jhun2009boot}. Moreover, the LD estimator \eqref{ld_reg} has the desired theoretical properties such as consistency and asymptotic normality \citep{Bai1990lde} as the conventional LS estimator \eqref{linear_reg} does.
A number of recent studies have proposed a robust regression including a deep robust regression model in a multiple regression problem \citep{diskin2017deep}, and robust DNN regression for a functional data \citep{wang2022robust}. However, nothing has been addressed in the multivariate regression problem setting.

\begin{figure}[t]
        \centering
        \includegraphics[width=.7\linewidth]{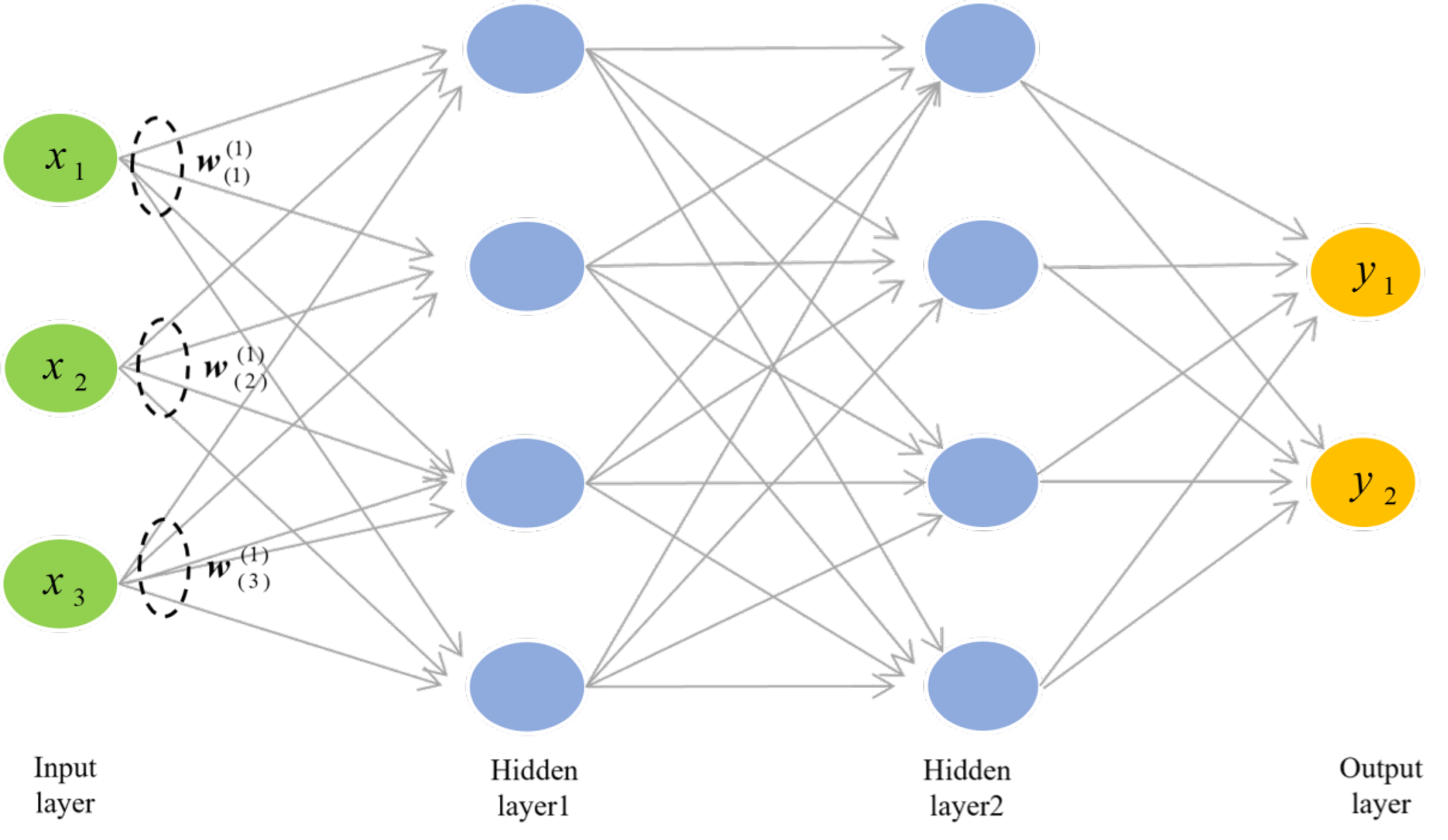}  
        \caption{An example of how group Lasso penalization works in the fully connected DNN architecture of the two hidden layers. Weights between the first and the second layer of rounded with black dashed line indicate the group of the weights to be shrunk towards zero vector \citep{Ho2020}.}
       \label{fig:GDNN_ARCH}
\end{figure}

In a variety of modern applications, the predictor is often high-dimensional. In such high-dimensional data analysis, variable selection is crucial to train a model since uninformative (or weakly informative) predictors deteriorate the prediction performance of the model. A large number of researches have done regards to the variable selection and the penalization becomes canonical after the least absolute shrinkage and selection operator \citep[Lasso,][]{Tibshirani1996} proposed. The penalization can be employed to the neural network framework for variable selection. However, the conventional Lasso that penalizes the parameters separately may not be ideal in neural network-based regression models, since a variable cannot be removed until all connection has been shrunk to zero. To tackle this, \cite{zhao2015heterogeneous}, and \cite{scardapane2017group} suggested the group Lasso \citep{Lu2006model} and its adaptive version \citep{Wang2008} for variable selection in the neural network framework, respectively. Fig.\ref{fig:GDNN_ARCH} describes how a group-wise penalization to the first layer weights activates for a variable selection purpose in a neural network-based multivariate regression problem. Earlier works, however, focused on the LS estimation. 

In this paper, we propose the least-distance-based (Adaptive) Group Lasso penalized estimators in a (deep) neural network framework (hereafter, (A)GDNN-LD) for a multivariate nonlinear regression problem. Although we assume shallow fully connected neural networks in order to deliver the proposed ideas in a clear way, the proposed method can readily be extended to deeper neural network architectures, such as multi-layers perceptron (MLP) with more hidden layers. 

 The rest of the paper is organized as follows. Section \ref{sec2} introduces the LD loss function applied to the DNN-based multivariate nonlinear regression model. In Section \ref{sec3}, we propose to apply the group Lasso penalty to achieve the variable selection under the proposed model. A set of simulation study is conducted in Section \ref{sec4} demonstrating the promising performance of the proposed method, and real data application to the concrete slump test data is provided in Section \ref{sec5}. Some concluding remarks follow in Section \ref{sec6}.

\section{Deep neural network based multivariate nonlinear regression estimator}\label{sec2}
\begin{figure}[ht]
    \centering
    \includegraphics[width=0.8\linewidth]{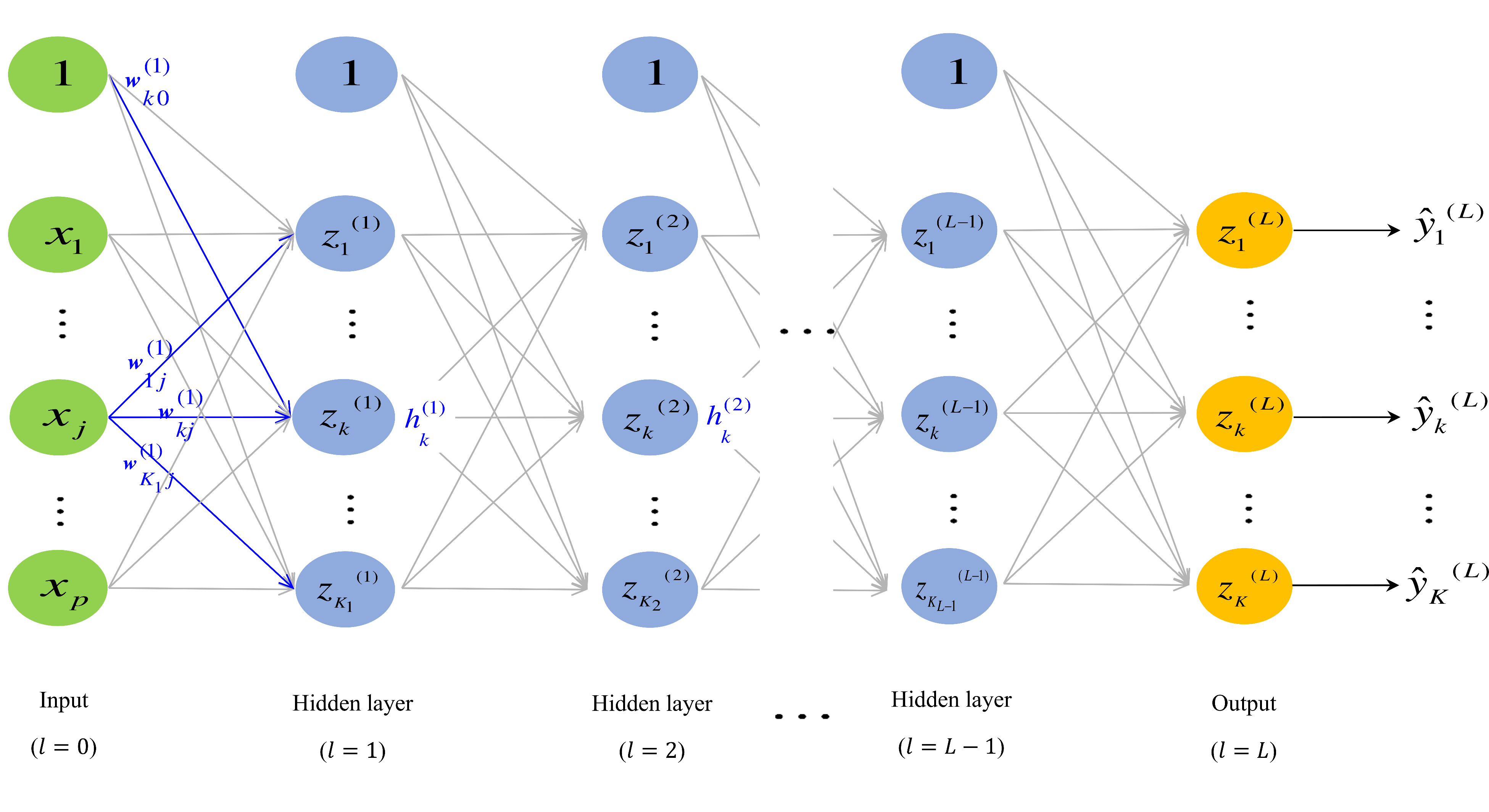}  
    \caption{Fully connected neural network architecture for a multivariate nonlinear regression with aforementioned notations. All bias nodes are set to one.}
    \label{fig:architecture.jpg}
\end{figure}

Given a set of data, $\left\{\left(\bx_{i}, \by_i \right) \right\}_{i=1}^n$, we want to estimate a multivariate nonlinear regression function $\boldsymbol{f(\bx)}$ in \eqref{model} assumed to have a deep neural network structure. As shown in Fig. \ref{fig:architecture.jpg}, we primarily consider a fully connected deep neural network with $(L+1)$ layers. The $l$ represents an order of layers from 0 to $L$. Let the $l=0$ layer be an input layer, the $l=L$ layer be an output layer, and the other layers are hidden layers. In this network, the $l$th layer consists of $(K_l + 1)$ nodes including a bias node, which has a common value of one, and we have $K_0 = p+1$ in the input layer and $K_L = q$ in the output layer.
We denote the $K_l \times (K_{l-1} + 1)$ weight matrix $\bW^{(l)}$ which connects the $(l-1)$th layer to the $l$th layer as below:
\begin{equation}\label{weight_mat}
 \bW^{(l)} = \left[ \bw_1^{(l)} \cdots \bw_k^{(l)} \cdots \bw_{K_l}^{(l)}\right]^{\top} = \left[ \bw_{(0)}^{(l)} \bw_{(1)}^{(l)} \cdots \bw_{(k)}^{(l)} \cdots \bw_{(K_{l-1})}^{(l)}\right],
\end{equation}
where $\bw_k^{(l)}=\left(w_{k 0}^{(l)}, \ldots, w_{k K_{l-1}}^{(l)}\right)^T$ represents the weight vector that connects the $(l-1)$th layer to $k$th node in the $l$th layer, and alternatively, $\bw_{(k)}^{(l)}=\left(w_{1 k}^{(l)}, \ldots, w_{K_l k}^{(l)}\right)^T$ represents the weight vector that links between $k$th node in the $(l-1)$th layer and the $l$th layer. The input data is fed in the forward direction from the input layer to the output layer in the following fashion,
$$
   \boldsymbol{z}^{(l)}=\bW^{(l)} \mathbf{h}^{(l-1)}, \quad \mathbf{h}^{(l)}=\phi_l\left(\boldsymbol{z}^{(l)}\right) \quad \text{for} \quad l=1, \cdots, L ,
$$
where $\mathbf{h}^{(l-1)}=\left(1, h_1^{(l-1)}, \cdots, h_{K_{l-1}}^{(l-1)}\right)^T$ with $h_k^{(l)}=\phi_l\left(z_k^{(l)}\right)$, for $k=1, \cdots, K_l$ is the feature vector in the $(l-1)$th layer,  $\boldsymbol{z}^{(l)}=\left(z_1^{(l)}, \cdots, z_{K_l}^{(l)}\right)^T$  with $z_k^{(l)}=\mathbf{w}_k^{(l)} h_k^{(l-1)}$, and $\phi_l (\cdot)$ is an activation function applied to the $l$th layer. 
The feature vector in the output layer $\hat{\by}^{(L)}=\mathbf{h}^{(L)}=\boldsymbol{z}^{(L)}=\bW^{(L)} \mathbf{h}^{(L-1)} \in \mathbb{R}^{K_L}$, which means the estimated value of the responses, is calculated with the identity activation function.

According to the universal approximation theorem, the DNN structure with appropriate number of hidden layers can approximate any arbitrary complex and continuous functions in a real space \citep[]{Cybenko1989, HORNIK1991}. Therefore, a neural network architecture can be a convenient tool for multivariate nonlinear regression analysis, and there have been several studies of the DNN-based least squares (DNN-LS) estimator in a multivariate regression problem setting \citep[]{Jiang2022, Zhang2019, Ochiai2017, Feng2017}.

The DNN-LS estimates a set of all weight matrices of the neural networks, $\calW=\left\{\bW^{(1)}, \cdots, \bW^{(L)}\right\}$ in \eqref{dnn-ls}
\begin{equation}\label{dnn-ls}
\widehat{\cal{W}}^{\textrm{DNN-LS}}=\underset{\bW^{(l)} \in \calW}{\arg \min } \sum_{i=1}^n\left\|\boldsymbol{y}_i-\bW^{(L)} \mathbf{h}_i^{(L-1)}\right\|^2,
\end{equation}
where $\widehat{\cal{W}}^{\textrm{DNN-LS}}=\left\{\widehat{\bW}^{(1), \textrm{DNN-LS}}, \cdots, \widehat{\bW}^{(L), \textrm{DNN-LS}}\right\}$, which is a set of estimated weight matrices with $\bW^{(l)} \in \mathbb{R}^{K_l \times (K_{l-1} + 1)}$ for $l=1,\cdots,L$, and $\mathbf{h}_i^{(L-1)}$ is the feature vector of $i$th observation in the $(L-1)$th hidden layer, which is defined as \eqref{feedforward_composite}.
\begin{equation}\label{feedforward_composite}
\mathbf{h}_i^{(L-1)}=\phi_{L-1}\left(\bW^{(L-1)}\left(\phi_{L-2}\left(\bW^{(L-2)}\left(\cdots \phi_2\left(\bW^{(2)}\left(\phi_1\left(\bW^{(1)} \bx_i\right)\right)\right)\right)\right)\right)\right),
\end{equation}
where $\phi_i(\cdot)$, $i=1,2,\cdots,L-1$ is an activation function for each layer. 
Then, the set of $\calW$, is updated via backpropagation \citep{Rumelhart1986}.
For example, equation \eqref{dnn_ls_update} shows how the weight vector $\bw_k^{(L)}$ defined in \eqref{weight_mat}, which connects the $(L-1)$th hidden layer to the $k$th node in the last hidden layer, is updated.
\begin{align}\label{dnn_ls_update}
\bw_k^{(L)*} &\leftarrow \bw_k^{(L)} - \eta  \frac{\partial J(\bw)}{\partial \bw_k^{(L)}},
\end{align}
where $\bw_k^{(L)*}$ is a newly updated weight for $k=1,\cdots,K_L$, $J(\bw)$ refers the cost function in \eqref{dnn-ls}, and $\eta$ denotes a learning rate. By the chain rule, the gradient of the loss function with respect to $\bw_k^{(L)}$ can be computed during the backpropagation as follows:
\begin{align} \label{ls_chain_rule}
\frac{\partial J(\bw)}{\partial \bw_k^{(L)}}=\frac{\partial J(\bw)}{\partial \hat{\by}_k^{(L)}} \frac{\partial \hat{\by}_k^{(L)}}{\partial {\bz_k}^{(L)}} \frac{\partial {\bz_k}^{(L)}}{\partial \bw_k^{(L)}},
\end{align}
and thus \eqref{dnn_ls_update} is
\begin{equation} \label{back_prop}
\bw_k^{(L)*} \leftarrow \bw_k^{(L)} + 2 \eta \left(\by_{k} - \hat{\by}_{k}^{(L)}\right) \phi^{\prime}(\bz_{k}^{(L)}) \boldsymbol{h}^{(L-1)},    
\end{equation}
where $\hat{\by}_k^{(L)}$ is the estimate of the $k$th response. We note from \eqref{back_prop} that only the residual of the $k$th response is used for updating $\bw_k^{(L)}$.

Similarly, $\bw_{k}^{(L-1)}$ is updated as
\begin{align}\label{dnn_ls_update_pen}
\bw_k^{(L-1)*} &\leftarrow \bw_k^{(L-1)} - \eta \frac{\partial J(\bw)}{\partial \bw_k^{(L-1)}}.
\end{align}
where the gradient $\frac{\partial J(\bw)}{\partial \bw_k^{(L-1)}}$ is given by
\begin{align}\label{dnn_ls_update_derv}
\frac{\partial J(\bw)}{\partial \bw_k^{(L-1)}}  &= \frac{\partial J(\bw)}{\partial \hat{\mathbf{h}}_k^{(L-1)}}
\frac{\partial \hat{\mathbf{h}}_k^{(L-1)}}{\partial {\bz_k}^{(L-1)}} \frac{\partial {\bz_k}^{(L-1)}}{\partial \bw_k^{(L-1)}} \\
&=\left\{ \sum_{k=1}^{K_{L}} \frac{\partial \left(\by_k - \hat{\by}_k^{(L)} \right)^2}{\partial \hat{\by}_k^{(L)}} \frac{\partial \hat{\by}_k^{(L)}}{\partial \bz_k^{(L)}} \frac{\partial \bz_k^{(L)}}{\partial \mathbf{h}_k^{(L-1)}} \right\} \frac{\partial \hat{\mathbf{h}}_k^{(L-1)}}{\partial {\bz_k}^{(L-1)}} \frac{\partial {\bz_k}^{(L-1)}}{\partial \bw_k^{(L-1)}} \\
&= -2\left\{ \sum_{k=1}^{K_{L}} \left(\by_k - \hat{\by}_k^{(L)}\right)\phi^{\prime}(\bz_{k}^{(L)})\bw_{k}^{(L)} \right\}\phi^{\prime}(\bz_{k}^{(L-1)})\mathbf{h}_k^{(L-2)}. \label{last_in_15}
\end{align}

It is important to note from (\ref{last_in_15}) that all residuals associated with every response variable including the $k$th one are calculated when updating $\bw_{k}^{(L-1)}$. This implies that the dependency structure among all responses is used for updating the weight matrix $\bW^{(L-1)}$. Furthermore, this logic is applied identically to the rest of weight parameters to update, i.e., $\left(\bW^{(1)*}, \cdots, \bW^{(L-2)*}\right)$. However, the DNN-LS estimator cannot use interdependent information of the correlated response variables when updating $\bW^{(L)}$ the weight parameter associated with the last layer.

\begin{figure}[t]
    \subfloat[ Compare least squares loss and least distance loss by their shapes.]{{\includegraphics[width=6cm]{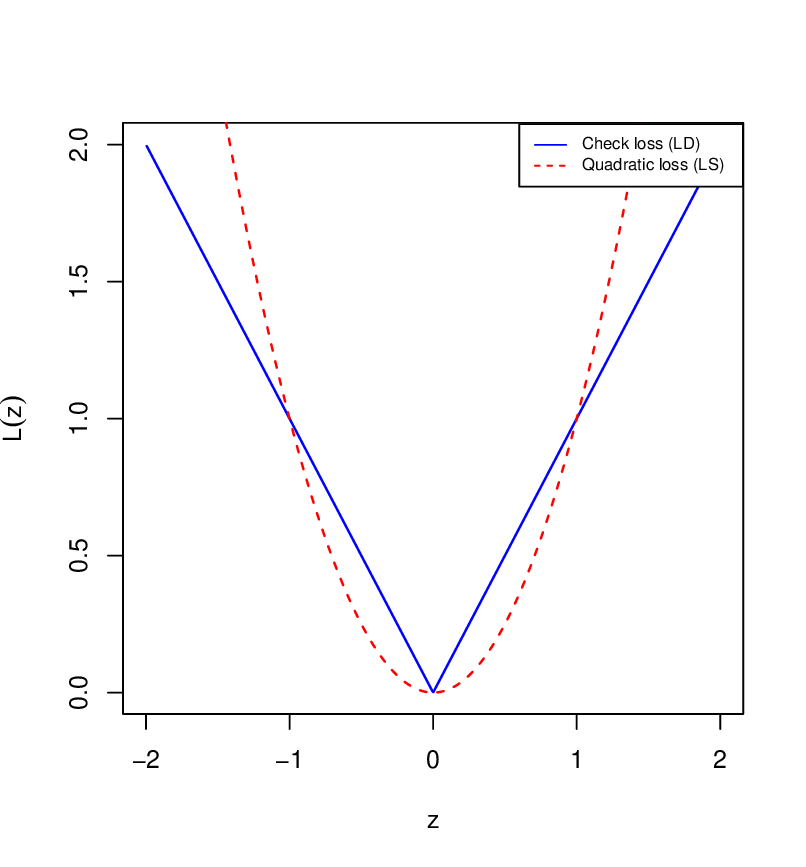} }}%
    \qquad
    \subfloat[ The quadratic smoothing functions are displayed with its original check function shape with $\tau=0.3$.]{{\includegraphics[width=6cm]{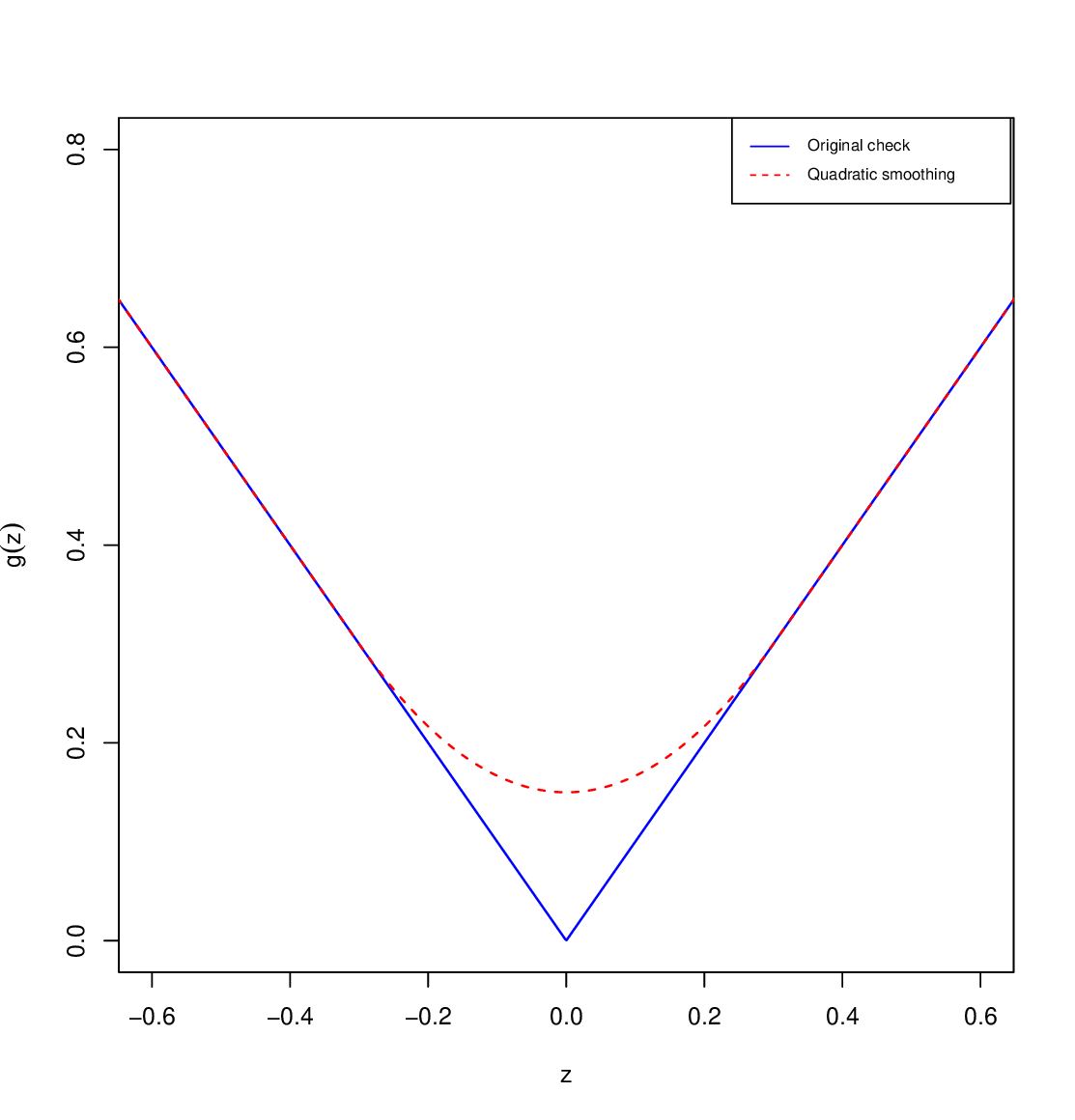} }}%
    \caption{The left panel is for viewing the characteristic of LD loss over LS loss and the right panel shows the quadratic smoothing approximation function suggested in \eqref{smoothing}.}
    \label{fig:Fig.2}
\end{figure}

To overcome such limitations, we propose the DNN-based Least Distance (DNN-LD) estimator, which estimates the weight matrices by applying the least distance loss function to the existing fully connected DNN architecture as formulated below \eqref{DNN-LD}.
\begin{equation}\label{DNN-LD}
\widehat{\cal{W}}^{\textrm{DNN-LD}}=\underset{\bW^{(l)} \in \calW}{\arg \min } \frac{1}{n} \sum_{i=1}^n\left\|\boldsymbol{y}_i-\bW^{(L)} \mathbf{h}_i^{(L-1)}\right\|_2,
\end{equation}
where $\widehat{\cal{W}}^{\textrm{DNN-LD}}=\left\{\widehat{\bW}^{(1), \textrm{DNN-LD}}, \cdots, \widehat{\bW}^{(L), \textrm{DNN-LD}}\right\}$, which is a set of estimated weight matrices.
Given a set of initial weight matrices $\calW^0=\left\{\bW^{(1),0}, \cdots, \bW^{(L),0}\right\}$, we can update the weights iteratively via gradient descent as well. The following \eqref{dnn_ld_update} shows how $\bw_k^{(L)}$ is updated:

\begin{align}\label{dnn_ld_update}
\bw_k^{(L)*} &\leftarrow \bw_k^{(L)} - \eta \frac{\partial J_{\textrm{LD}}(\bw)}{\partial \bw_k^{(L)}},
\end{align}
where for $k=1,\cdots,K_L$ and the cost function in \eqref{DNN-LD} denotes as $J_{\textrm{LD}}(\bw)$.
The gradient $\frac{\partial  J_{\textrm{LD}}(\bw)}{\partial \bw_k^{(L)}}$ in \eqref{dnn_ld_update} can be reformulated as \eqref{dnn_ld_update_2}-\eqref{dnn_ld_update_3} by the chain rule.
\begin{align}\label{dnn_ld_update_2}
\frac{\partial J_{\textrm{LD}}(\bw)}{\partial \bw_k^{(L)}}  &= \frac{\partial J_{\textrm{LD}}(\bw)}{\partial \hat{\by_k}^{(L)}} 
\frac{\partial \hat{\by_k}^{(L)}}{\partial {\bz_k}^{(L)}} \frac{\partial {\bz_k}^{(L)}}{\partial \bw_k^{(L)}} \\
&=-\frac{\left( y_k - \hat{y}_k^{(L)}\right)}{ \left\| \by - \hat{\by}^{(L)}\right\|_{2}}\phi^{\prime}(\bz_{k}^{(L)}) \boldsymbol{h}^{(L-1)}. \label{dnn_ld_update_3}
\end{align}
Hence, we can rewrite (\ref{dnn_ld_update}) as \eqref{back_prop_ld}:
\begin{equation} \label{back_prop_ld}
\bw_k^{(L)*} \leftarrow \bw_k^{(L)} + \eta \frac{\left( y_k - \hat{y}_k^{(L)}\right)}{ \left\| \by - \hat{\by}^{(L)}\right\|_{2}} \phi^{\prime}(\bz_{k}^{(L)}) \boldsymbol{h}^{(L-1)}.   
\end{equation}
As shown in \eqref{back_prop_ld}, the DNN-LD is able to exploit the residuals of all responses by the form of $\left\| \cdot \right\|_2$ even when updating the $\bw_k^{(L)}$, and thus $\bW^{(L)}$. It is thus not difficult to see that the DNN-LD estimator will show better performance as response variables are correlated more. This point will be revealed by simulations in Section \ref{sec3}. 

We note that \eqref{DNN-LD} is non-differentiable at the origin, which makes its optimization non-trivial. To address this difficulty, the smooth approximation proposed by \citet{bian2012smoothing} is applied to the original absolute loss function. The smoothed loss \eqref{smoothing} borrows the idea of Huberization \citep{huber1992robust} and smoothly transits the absolute loss to the squared loss at the $\tau$-neighborhood of the  origin to ensure differentiability. Here, $\tau$ is a positive constant working as a hyperparameter. As $\tau \rightarrow 0$, the smoothed loss function converges to the absolute loss function. Fig. \ref{fig:Fig.2} (b) shows the smoothing function with the original absolute loss function. 
\begin{align}\label{smoothing}
\mathrm{g}(\boldsymbol{z})= \begin{cases}\|\boldsymbol{z}\|, & (\|\boldsymbol{z}\| \geq \tau) \\ \frac{\|\boldsymbol{z}\|^2}{2 \tau}+\frac{\tau}{2}, & (\|\boldsymbol{z}\|<\tau).\end{cases}
\end{align}

If we reformulate $\frac{\partial J_{\textrm{LD}}(\bw)}{\partial \bw_k^{(L)}}$ in \eqref{dnn_ld_update_2} by applying \eqref{smoothing}  to the original cost function \eqref{DNN-LD}, we will get \eqref{smooth_update_ld},
\begin{align} \label{smooth_update_ld}
    \frac{\partial J_{\textrm{LD}}(\bw)}{\partial \bw_k^{(L)}} =\begin{cases}
            -\frac{\left( y_k - \hat{y}_k^{(L)}\right)}{ \left\| \by - \hat{\by}^{(L)}\right\|_{2}}\phi^{\prime}(\bz_{k}^{(L)}) \boldsymbol{h}^{(L-1)} & ,\left( \left\|\boldsymbol{y}-\hat{\by}^{(L)}\right\|_2 \geq \tau \right) \\
            -\frac{1}{\tau} \left\|\boldsymbol{y}-\hat{\by}^{(L)}\right\|_{2} \left(y_k-y_k^{(L)}\right) \phi^{\prime}(\bz_{k}^{(L)}) \boldsymbol{h}^{(L-1)}&
            ,\left( \left\|\boldsymbol{y}-\hat{\by}^{(L)}\right\|_2 < \tau \right).
            \end{cases}
\end{align}
Note that even after applying the approximation \eqref{smoothing}, the gradient still contains norms of all residual vectors and thus the corresponding estimator possesses the advantages of the DNN-LD with the original loss function. The estimation process of the smoothed DNN-LD method is summarized in Algorithm \ref{alg:smoothing}.
\begin{algorithm}[h]
    \caption{Backpropagation for DNN-LD with smoothing approximation}\label{alg:smoothing}
    \begin{algorithmic}[1]
    \For {l=1 to  L-1} :
        \For{k=0 to $K_l$ }
            \State $\delta_k^{(l)}=-e_k^{(l)} \phi_l^{\prime}\left(z_k^{(l)}\right),$
            \State $\quad \text{where} \quad e_k^{(l)}=\boldsymbol{w}_{(k)}^{(l+1)^T} \boldsymbol{\delta}^{(l+1)}, \quad \boldsymbol{\delta}^{(l+1)}=\left(\delta_1^{(l+1)}, \ldots, \delta_{K_l}^{(l+1)}\right)^T. $
            \State $\boldsymbol{w}_k^{(l)} \leftarrow \boldsymbol{w}_k^{(l)}-\eta \delta_k^{(l)} \mathbf{h}^{(l-1)}$ \Comment{applied gradient descent method}
        \EndFor
    \EndFor
    \For {l=L} :
        \For{k=1 to q } :
            \State \small{$\delta_k^{(L)}=\begin{cases}
            -\frac{\left(y_k-h_k^{(L)}\right)}{\left\|\boldsymbol{y}-\mathbf{h}^{(L)}\right\|_2} & ,\left( \left\|\boldsymbol{y}-\mathbf{h}^{(L)}\right\|_2 \geq \tau \right) \\
            -\frac{1}{\tau} \left\|\boldsymbol{y}-\mathbf{h}^{(L)}\right\|_2 \left(y_k-h_k^{(L)}\right) &
            ,\left( \left\|\boldsymbol{y}-\mathbf{h}^{(L)}\right\|_2 < \tau \right)
            \end{cases}$} \\\Comment{applied smoothing method}
            \State$\boldsymbol{w}_k^{(L)} \leftarrow \boldsymbol{w}_k^{(L)}-\eta \delta_k^{(L)} \mathbf{h}^{(L-1)}$ 
        \EndFor
    \EndFor
    \end{algorithmic}
\end{algorithm}

\section{Adaptive group Lasso penalized DNN-LD estimator for variable selection} \label{sec3}
With a high-dimensional predictor, the variable selection is an important task. Toward this, we propose to employ the group Lasso penalty \citep{Lu2006model} to the DNN-LD estimator.
Given $n$ independent copies of $(\bx, \by)$, denoted by $\left\{\left(\bx_i, \boldsymbol{y}_i\right)\right\}_{i=1}^n$, the group Lasso DNN-LD (GDNN-LD) solves
\begin{equation}\label{eq:GDNN-LD}
\widehat{\cal{W}}^{\textrm{GDNN-LD}}=\underset{\bW^{(l)} \in \calW}{\arg \min } \frac{1}{n} \sum_{i=1}^n\left\|\by_i-\bW^{(L)} \mathbf{h}_i^{(L-1)}\right\|+\lambda \sum_{j=0}^p\left\|\bw_{(j)}^{(1)}\right\|,
\end{equation}
where $\bw_{(j)}^{(1)}=\left(w_{1 j}^{(1)}, w_{2 j}^{(1)}, \ldots, w_{K_1 j}^{(1)}\right)^T$, which means the weight vector links $j$th explanatory variable $x_j$ to the first hidden layer nodes. Here, $\lambda$ is a hyperparameter which modulates the magnitude of the penalty term, which will be decided from the data. The GDNN-LD imposes the penalty group-wisely onto the weights $w_{k j}^{(1)} \left(k=0, \cdots, K_1, j=0,\cdots,p\right)$ by the importance of the predictor $x_j$ to predict the responses in the model in \eqref{DNN-LD}. Fig. \ref{fig:GDNN_ARCH} well represents the scheme of a variable selection in the DNN framework \citep{Ho2020}.  
However, it is well known that the Lasso and its variants such as group Lasso yield bias \citep[]{Fan2001selec, Wang2008} and the corresponding estimator could be inconsistent \citep[]{Lu2006model, Zou2006:Oral}. To overcome the drawback, \citet{Wang2008} suggested the adaptive group Lasso to mitigate the bias of the group Lasso. 
This motivates us to propose the adaptive group Lasso DNN least distance method (AGDNN-LD) as follows:
\begin{equation}\label{AGDNN-LD}
\widehat{\cal{W}}^{\textrm{AGDNN-LD}}=\underset{\bW^{(l)} \in \calW}{\arg \min } \frac{1}{n} \sum_{i=1}^n\left\|\by_i-\bW^L \mathbf{h}_i^{(L-1)}\right\|+\sum_{j=0}^p \lambda_j^{\textrm{DNN-LD}}\left\|\bw_{(j)}^{(1)}\right\|,
\end{equation}
where $\lambda_j^{\textrm{DNN-LD}}=\lambda\left(\widehat{\bw}_{(j)}^{(1),\textrm{DNN-LD}}\right)^{-\gamma} \in \mathbb{R}, (j=0,1, \cdots, p), \lambda > 0$. Here, $\widehat{\bw}_{(j)}^{(1),\textrm{DNN-LD}}$ means the estimated weight vector by the DNN-LD. Also, $\gamma$ is a positive constant, which is not overly sensitive. So, it is set to 1 in this paper for simplicity. The adaptive Lasso enjoys better shrinkage performance by varying the degree of penalization to each group by reducing the bias in the variable selection procedure.

 Meanwhile, the penalty term for both \eqref{eq:GDNN-LD} and \eqref{AGDNN-LD}, also suffer a non-differentiability problem. To gain a theoretical completeness, we borrow the quadratic smoothing function given in \eqref{smoothing} again. However, the term of $\left\|\bw_{(j)}^{(1)}\right\|$ is unable to be shrunk to exact zero vector under the approximation, we will consider the $j$th  predictor $x_j$ as an irrelevant variable when $\left\| \widehat{\bw}_{(j)}^{(1)}\right\|^2 \le 10^{-3}$ for $j=0,\cdots,p$ with enoughly small $\tau=10^{-5}$.
The backpropagation process for the GDNN-LD is summarized in Algorithm \ref{alg:GL_smoothing}.

\begin{algorithm}[t]
    \caption{Backpropagation for GDNN-LD with smoothing approximation}\label{alg:GL_smoothing}
    \begin{algorithmic}[1]
    \For {l=1 } :
        \For{k=0 to $p$ }
            \State $\delta_k^{(l)}=-e_k^{(l)} \phi_l^{\prime}\left(z_k^{(l)}\right),$
            \State $\quad \text{where} \quad e_k^{(l)}=\boldsymbol{w}_{(k)}^{(l+1)^T} \boldsymbol{\delta}^{(l+1)}, \quad \boldsymbol{\delta}^{(l+1)}=\left(\delta_1^{(l+1)}, \ldots, \delta_{K_l}^{(l+1)}\right)^T. $
            \State \small{$\boldsymbol{p}_k^{(1)} = \begin{cases} \lambda \frac{\boldsymbol{w}_{(k)}^{(1)}}{\left\|\boldsymbol{w}_{(k)}^{(1)}\right\|}, & \left( \left\|\boldsymbol{w}_{(k)}^{(1)}\right\| \geq \tau_1 \right) \\
            \frac{\lambda}{\tau_1}\boldsymbol{w}_{(k)}^{(1)}\left\|\boldsymbol{w}_{(k)}^{(1)}\right\|, &
            \left( \left\|\boldsymbol{w}_{(k)}^{(1)}\right\| < \tau_1 \right)
            \end{cases}$} \Comment{applied smoothing method}\\
            \Comment{add Group Lasso penalty to the loss only for the first layer}
            \State $\boldsymbol{w}_k^{(l)} \leftarrow \boldsymbol{w}_k^{(l)}-\eta \left[ \delta_k^{(l)} \mathbf{h}^{(l-1)} + \boldsymbol{p}_k^{(1)} \right] $
        \EndFor
    \EndFor
    
    \For {l=2 to  L-1} :
        \For{k=0 to $K_l$ }
            \State $\delta_k^{(l)}=-e_k^{(l)} \phi_l^{\prime}\left(z_k^{(l)}\right),$
            \State $\quad \text{where} \quad e_k^{(l)}=\boldsymbol{w}_{(k)}^{(l+1)^T} \boldsymbol{\delta}^{(l+1)}, \quad \boldsymbol{\delta}^{(l+1)}=\left(\delta_1^{(l+1)}, \ldots, \delta_{K_l}^{(l+1)}\right)^T. $
            \State $\boldsymbol{w}_k^{(l)} \leftarrow \boldsymbol{w}_k^{(l)}-\eta  \delta_k^{(l)} \mathbf{h}^{(l-1)} $
        \EndFor
    \EndFor
    \For {l=L} :
        \For{k=1 to q } :
            \State $\delta_k^{(L)}=\begin{cases}
            -\frac{\left(y_k-h_k^{(L)}\right)}{\left\|\boldsymbol{y}-\mathbf{h}^{(L)}\right\|}, & \left( \left\|\boldsymbol{y}-\mathbf{h}^{(L)}\right\| \geq \tau_2 \right) \\
            -\frac{1}{\tau_2} \left\|\boldsymbol{y}-\mathbf{h}^{(L)}\right\| \left(y_k-h_k^{(L)}\right), &
            \left( \left\|\boldsymbol{y}-\mathbf{h}^{(L)}\right\| < \tau_2 \right)
            \end{cases}$ \\ 
            \State $\boldsymbol{w}_k^{(L)} \leftarrow \boldsymbol{w}_k^{(L)} - \eta  \left\{ \delta_k^{(L)} \mathbf{h}^{(L-1)} \right\} $
        \EndFor
    \EndFor
    \end{algorithmic}
\end{algorithm}

\section{Simulations}\label{sec4}
In this section, we perform a couple of simulations to verify the efficiencies of the proposed methods. Example 1 is designed to show the prediction performance of the DNN-LD estimator in a nonlinear multivariate regression setup by changing correlation levels of the response variables. In Example 2, we consider a situation where outliers are present in the data to show the robustness of the proposed DNN-LD estimator. Finally, in Example 3, we show a variable selection property of the  (A)GDNN-LD estimator. Commonly, we generated 200 training and 200 validation observations, along with $n_{\text{test}}(=10,000)$  independent test observations. Then run each simulation 100 times to address the sampling variability. To estimate the model error, we calculated the mean squared error (MSE) on the test data, defined as
%$$
%\text{MSE}=\frac{1}{n_{\text {test }} q} \sum_{k=1}^q \sum_{i=1}^{n_{\text {test }}}\left\|y_{i k}-\hat{f}_k\left(\bx\right)\right\|^2.
%$$
In terms of initial parameters of the deep learning model, we used 2 hidden layers of 10 nodes each with sigmoid activation function, and a learning rate $\eta = 0.01 \sim 0.3$ is considered. Also, the cutoff value $\tau$ regarding to the quadratic smoothing function in \eqref{smoothing} is set to $10^{-3}$. 
As for a computing software, we implemented the DNN model with our own code in R. The R code is available upon request.
%These set of hyperparameters are the pre-determined from the grid search method for each simulation case. 
%Moreover, to avoid the overfitting issue, we administer a couple of remedies such as batch normalization \citep{Ioffe2015}, Adam optimizer \citep{Kingma2014:adam}, and Xavier initialization \citep{Glorot2010}.

\begin{figure}[t]
    \centering
    \subfloat[Graph of the three response variables, $y_1, y_2, y_3$ from case 1 in Example 1 \eqref{ex1-1}. The pairwise distance correlations of $(y_1,y_2),(y_1, y_3), (y_2, y_3)$ from the training data are 0.71, 0.62, 0.65, respectively.]{\includegraphics[width=6.0cm]{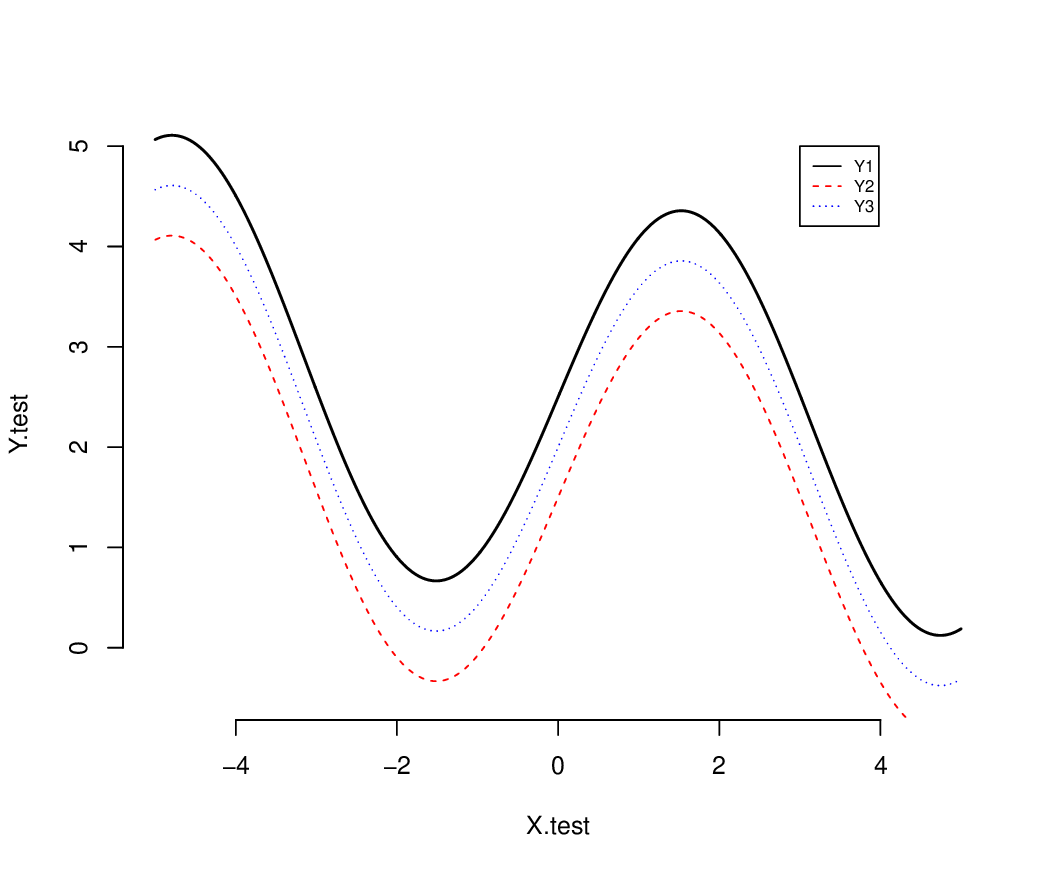} }%
    \quad
    \subfloat[The plots of case 2 in Example 1 \eqref{ex1-2} are also displayed. The pairwise distance correlations of the training data of $(y_1,y_2),(y_1, y_3), (y_2, y_3)$ are 0.17, 0.12, 0.21, respectively.]{{\includegraphics[width=6.0cm]{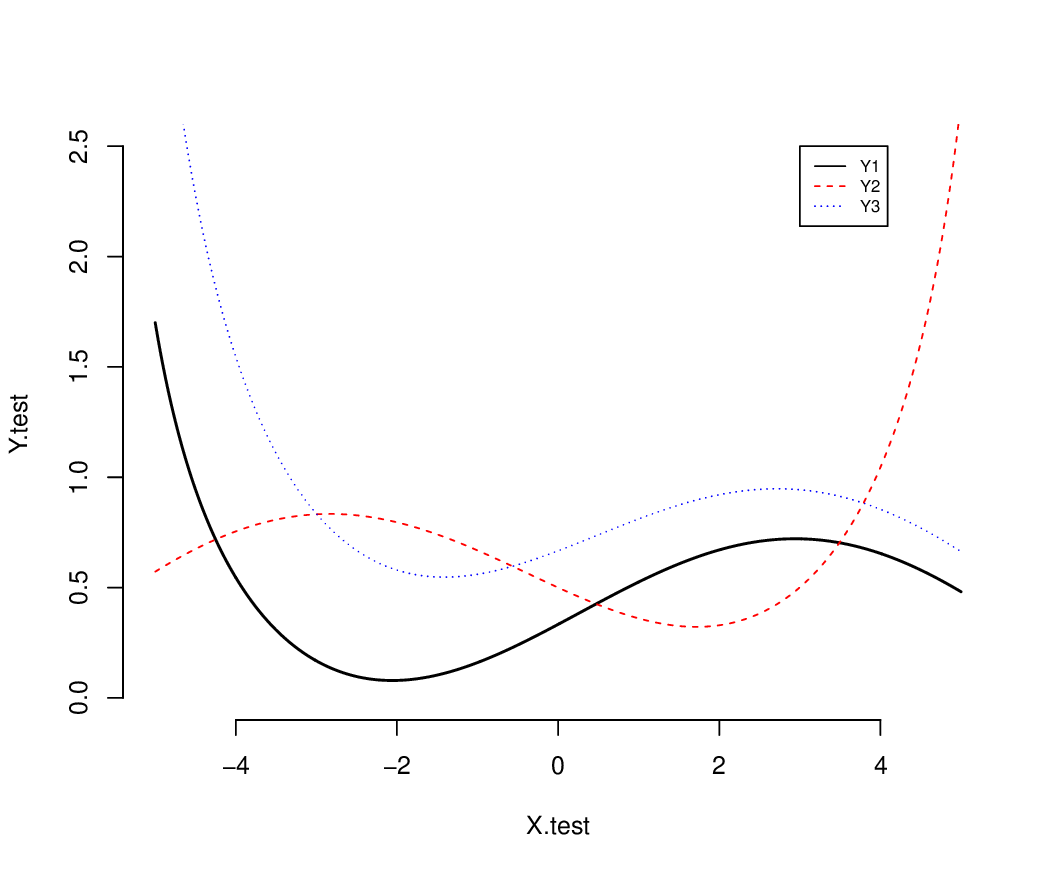} }}%
    \caption{The two true cases for Example 1 are displayed. The left panel supports that case 1 of a high level of correlation among the response variables, and the right panel, shows a weak relationship among the response variables.}
    \label{fig:ex_1_model}
\end{figure}

\subsection{Example 1}\label{sim:ex1}
%This example concerns the simultaneous estimation of multivariate nonlinear conditional mean functions with correlated multi responses using DNN architecture.
We generated the response vectors  $\by = (y_1, y_2, \cdots, y_q)^{\top}$ from the following two cases:

\begin{equation}\label{ex1-1}
    \text{Case 1} : y_k = \frac{k}{2} + 2\text{sin}(x_1) + \exp{(-0.1x_1)} + \epsilon_k,
\end{equation}

\begin{equation}\label{ex1-2}
    \text{Case 2} : y_k = \frac{(k+1)}{(-1)^{(k+1)} x_1 + 6} + \frac{1}{2}\text{sin}\left(\frac{(-1)^{(k+1)}}{2}x_1 \right)+ \epsilon_k,
\end{equation}
where for $k=1,\cdots,q$ ($q= 3, 6, 9$ are considered), and the predictor variable $x_1 \sim U(-5,5)$. A random error, $\epsilon_k$ is drawn from $N (0,1)$ and ${t}(3)$  independently. The two true cases are displayed in Fig. \ref{fig:ex_1_model}, and the responses are supposed to be correlated pairwisely, to some extent. %due to the similarity of $y_j$ for both case 1 and case 2. 
As for a correlation measure, we consider the distance correlation \citep[$dcor$]{Szekely2008} which can measure both linear and nonlinear association between two random vectors. The average pairwise distance correlations, $\overline{dcor}(y_k, y_{k^{\prime}})$, where $k, {k^{\prime}}=1, \cdots, q$ and $k \ne k^{\prime}$, of the two cases are  $0.7$ and  $0.2$, respectively.
Table \ref{tbl:ex1}  shows the average MSE for each estimation method. The kernel ridge regression (KRR), smoothing spline (SS), independently working DNN with least squares loss estimator (DNN-LS-IND), and DNN-LS are considered as competitive methods. The numbers given in parentheses are the standard errors.

\begin{center}
\begin{table}[t]
\resizebox{\textwidth}{!}{%
    \large{
    \begin{tabular}{ccccccccc}
    \hline
    Case        & $\overline{dcor}$          & $\epsilon$          & $q$ & SS & KRR & DNN-LS-IND & DNN-LS & DNN-LD \rule[0ex]{0pt}{3ex}\\ 
    \hline
    {1} & {0.7} & {$\mathrm{N(0,1)}$} & 3 & 0.067 (0.012) & 0.042 (0.016)  & 0.041 (0.014)   & \textbf{0.029} (0.013)  &  0.030 (0.014)  \rule[0ex]{0pt}{2ex}  \\
        &      &       & 6 & 0.067 (0.014) & 0.037 (0.006)   & 0.040 (0.019)   &0.028 (0.009)   & \textbf{0.027} (0.008)   \rule[0ex]{0pt}{2ex}    \\
        &      &       & 9 & 0.070 (0.014) & 0.041 (0.008)   & 0.049 (0.021)   &0.030 (0.006)   & \textbf{0.022} (0.005)   \rule[0ex]{0pt}{2ex}   \\ \cmidrule{3-9}
        &      & {$\mathrm{t(3)}$}  & 3 & 0.134 (0.026) & 0.114 (0.037)  & 0.137 (0.084)    & 0.078 (0.032)   & \textbf{0.046} (0.017)    \rule[0.5ex]{0pt}{2ex}  \\
        &      &       & 6 & 0.122 (0.027) & 0.103 (0.027)  & 0.098 (0.041)    & 0.060 (0.020)   & \textbf{0.033} (0.011)    \rule[0.5ex]{0pt}{2ex}  \\
        &      &       & 9 & 0.124 (0.022) & 0.106 (0.022)  & 0.086 (0.025)    & 0.058 (0.014)   & \textbf{0.030} (0.006)    \rule[0.5ex]{0pt}{2ex}  \\ \cmidrule{1-9}
    {2} & {0.2} & {$\mathrm{N(0,1)}$} & 3 & \textbf{0.035} (0.009) & 0.038 (0.008)  & 0.040 (0.017)   & 0.043 (0.016)   & 0.044 (0.014)    \rule[0.5ex]{0pt}{2ex}  \\
        &       &      & 6 & 0.032 (0.004) & 0.035 (0.011)  & 0.032 (0.021)   & 0.038 (0.008)   & \textbf{0.028} (0.005)    \rule[0.5ex]{0pt}{2ex}  \\
        &       &      & 9 & 0.042 (0.011) & 0.049 (0.014)  & 0.056 (0.025)   & 0.057 (0.013)   & \textbf{0.035} (0.008)    \rule[0.5ex]{0pt}{2ex}  \\\cmidrule{3-9}
        &       & {$\mathrm{t(3)}$} & 3 & 0.091 (0.032) & \textbf{0.076} (0.024)   &  0.093 (0.033)  & 0.114 (0.043)   & 0.093 (0.036)     \rule[0.5ex]{0pt}{2ex}  \\
        &       &      & 6 & 0.091 (0.015)  & 0.088 (0.021)   & \textbf{0.078} (0.049)  & 0.087 (0.022) & 0.085 (0.021)       \rule[0.5ex]{0pt}{2ex}  \\
        &       &      & 9 & 0.103 (0.033) & 0.108 (0.028)  & 0.100 (0.030)   & 0.093 (0.022)   & \textbf{0.086} (0.021)      \rule[0.5ex]{0pt}{2ex}  \\ \hline
    \end{tabular}%
    }}
    \caption{The averages and standard errors (in parentheses) of MSE based on 100 replications for SS, KRR, DNN-LS-IND, DNN-LS and DNN-LD for Example 1. The smallest MSE is written in bold.}
    \label{tbl:ex1}
\end{table}
\end{center}

\begin{figure}[t]
    \centering
    \subfloat[Example 1 result with $q=9$, \text{normal error} of case 1. The black solid line is the true $y_9$ and the red solid line is for the DNN-LD. The red solid line is the closest to the true line. The background points are training data.]{\includegraphics[width=6.0cm]{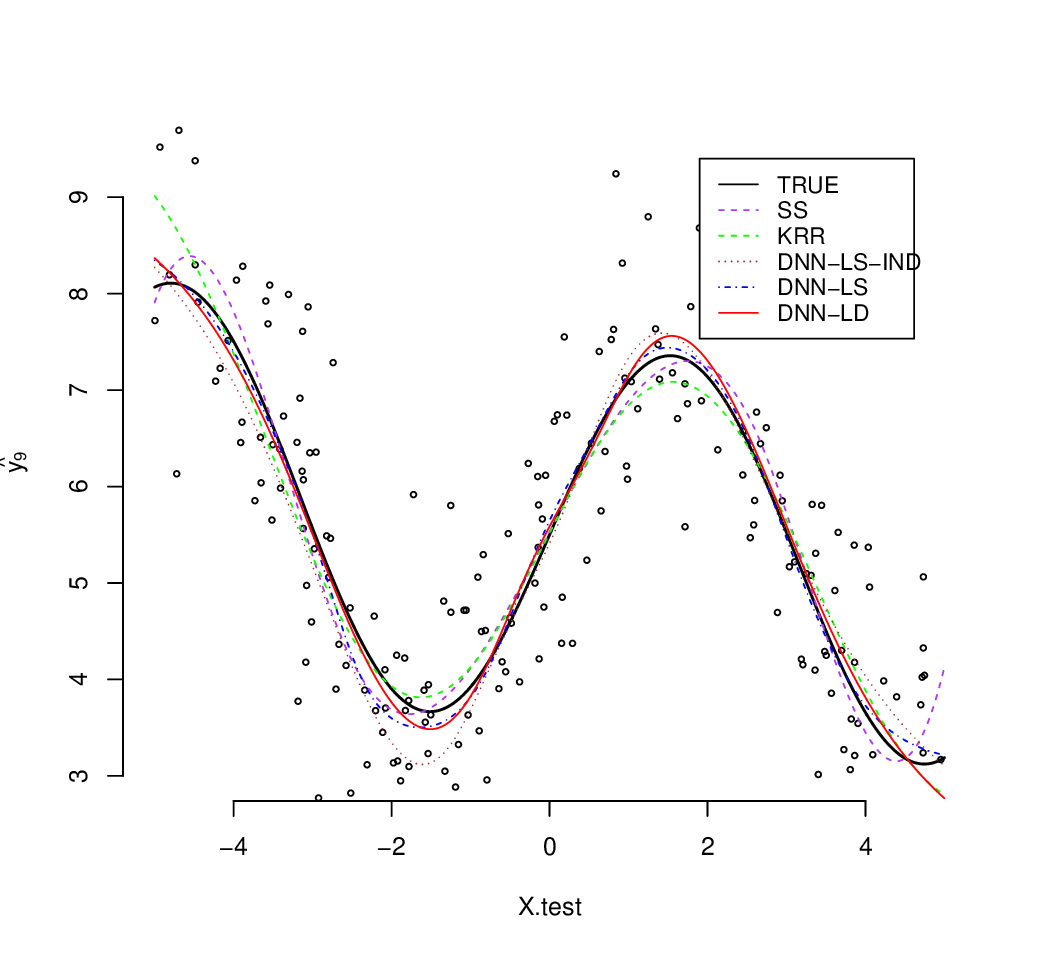} }%
    \quad
    \subfloat[The result with $q=9$, \text{t-distribution} of case 2. The black solid line is the true $y_9$ and the red solid line is for the DNN-LD estimator.]{\includegraphics[width=6.0cm]{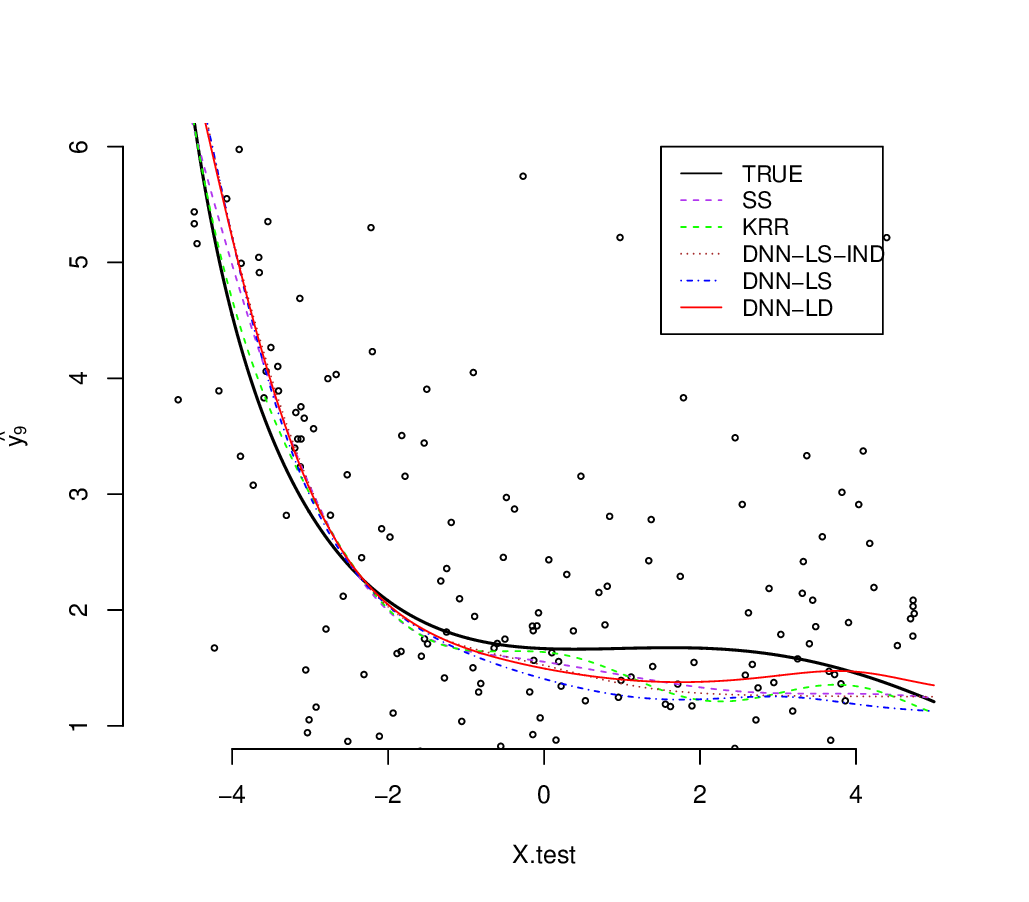}}%
    \caption{The two subfigures obtained from Example 1 visualize the prediction performance of the DNN-LD estimator over the competitors.}
    \label{fig:ex_1_result}
\end{figure}

The following conclusions can be drawn from Example 1. First, in the case of a high level of $\overline{dcor}$ among the responses, which is case 1, the DNN-LD methods tend to have better prediction performance than other competitors, and this tendency becomes more remarkable as $q$ increases. This tendency is more pronounced in case of having t-distribution as the random error distribution than normal distriubution. On the other hand, like case 2, with a weak distance correlation among responses, that is, each $y_k, k=1,2, \cdots, q$, which share little strength, the performance of the DNN-LD is relatively less distinct than the result of the first case. One $y_k$ may play as a noise for the other response, $y_{k^{\prime}}$. For this reason, when weak correlations exist, DNN-LD may not be a suitable choice, rather a individual estimation would be preferred. However, even in case 2, the DNN-LD shows a better prediction result then the DNN-LS. The related figures are presented in Fig.\ref{fig:ex_1_result}.
%Since, one $y_j$ could play as a noise for the other response, $y_k$.

\subsection{Example 2}\label{sim:ex2}
In this example, the robustness of the DNN-LD to the outliers, which is demonstrated by intentionally adding 5\%, 10\%, and 20\% of distorted points to the training data in the same simulation setting to case 1 in Example 1, is highlighted. In detail, add 7 to the odd-numbered responses of the training data as $y_{2k-1} + 7$ and subtract 7 from the even-numbered variables like $y_{2k} - 7$  $(k=1,\cdots,5)$ corresponding to the outlier rate $(\alpha)$ of 5\%, 10\% and 20\%. Both $q=3$ and $q=9$ cases are considered. Table \ref{tbl:ex2} summarizes the result of Example 2.

\begin{figure}[t]
    \subfloat[Example 2 result with $q=9, \alpha=20\%$ with $N(0,1)$ error distribution. The black solid line is the true $y_9$ and the red solid line is for the DNN-LD estimator. The red solid line is the closest to the true line.]{{\includegraphics[width=6cm]{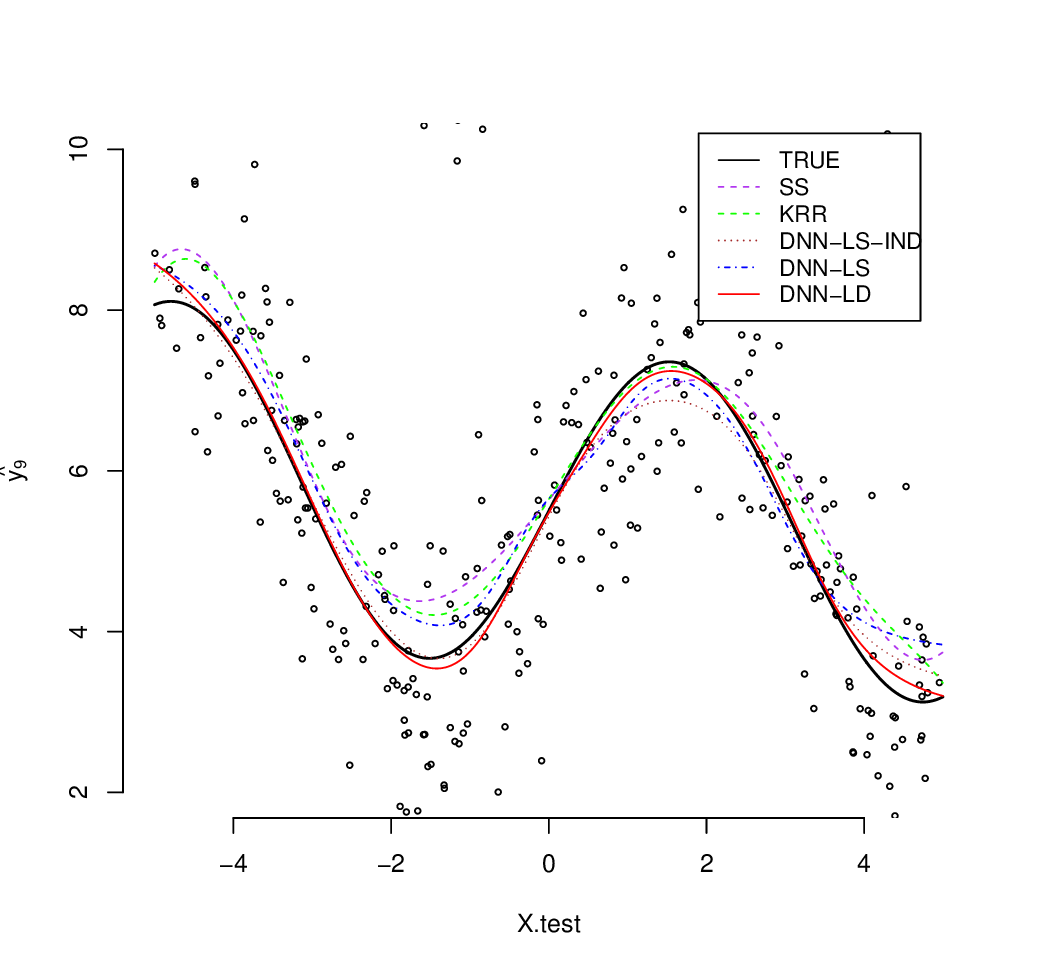} }}%
    \quad
    \subfloat[The result with $q=9, \alpha=20\%$ with $t(3)$ error distribution. The red solid line is for the DNN-LD estimator, which shows the best prediction performance.]{{\includegraphics[width=6cm]{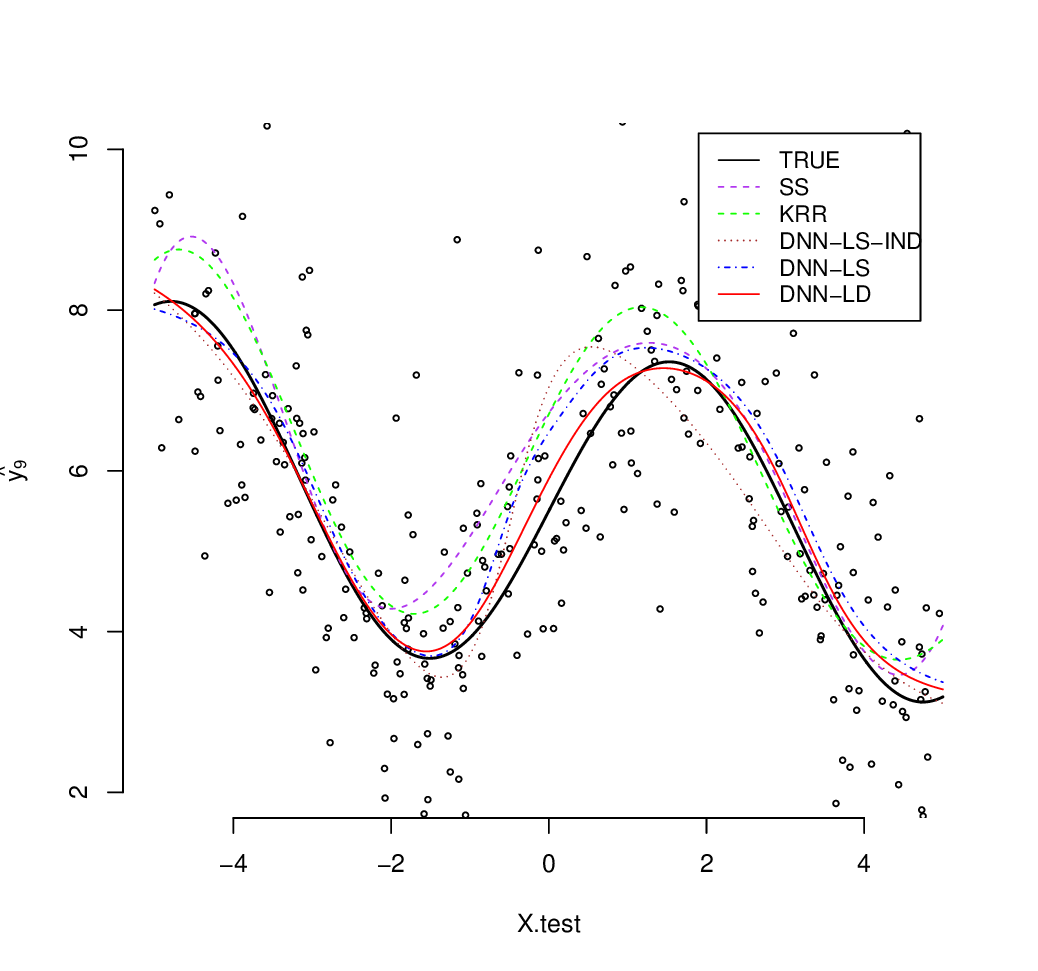} }}%
    \caption{The results of Example 2 is displayed. The two subfigures show the robustness of the DNN-LD estimator over the competitors. The background points are training data.}
    \label{fig:sim2}
\end{figure}

Both for the cases of $q=3$ and $q=9$, the DNN-LD method confirms the robustness to the outliers.%These results are held on the grounded that the characteristics of the least distance loss function and DNN architecture, which is available to incorporate correlation among the response variables into the estimation processes. 
Especially, in Table \ref{tbl:ex2}, the cases for both $q=3$ and $q=9$ with t distribution for the random error with 20\% of artificial outliers, the DNN-LD estimator yields much nicer results with respect to MSE compare to other estimation methods. In Fig.\ref{fig:sim2}, the two representative figures of the result are displayed.
\begin{table}[t]
\resizebox{\textwidth}{!}{%
\begin{tabular}{cccccccc}
\hline
$q$ & $\epsilon$  & $\alpha$    & SS            & KRR           & DNN-LS-IND    & DNN-LS        & DNN-LD  \rule[0ex]{0pt}{3ex} \\ \hline

3 & $\mathrm{N(0,1)}$ &  5\% & 0.077 (0.017) & 0.046 (0.016)  & 0.032 (0.018)  & 0.028 (0.009)  & \textbf{0.021} (0.007)  \rule[0.5ex]{0pt}{2ex}\\
  &    & 10\% & 0.153 (0.026) & 0.118 (0.028)  & 0.052 (0.024)  & 0.057 (0.015)  & \textbf{0.029} (0.008)  \rule[0.5ex]{0pt}{2ex}\\
  &    & 20\% & 0.327 (0.034) & 0.286 (0.033)  & 0.082 (0.053)  & 0.108 (0.026)  & \textbf{0.039} (0.012)  \rule[0.5ex]{0pt}{2ex}\\ \cmidrule{2-8} 
  & $\mathrm{t(3)}$ & 5\%  & 0.123 (0.039)  & 0.086 (0.032)   & 0.066 (0.025)  & 0.064 (0.024) & \textbf{0.030} (0.009)  \rule[0.5ex]{0pt}{2ex}\\
  &    & 10\% & 0.193 (0.030) & 0.152 (0.023)  & 0.118 (0.048)  & 0.096 (0.026)  & \textbf{0.044} (0.009)  \rule[0.5ex]{0pt}{2ex}\\
  &    & 20\% & 0.390 (0.048) & 0.340 (0.060)  & 0.129 (0.056)  & 0.187 (0.047)  & \textbf{0.065} (0.017)  \rule[0.5ex]{0pt}{2ex}\\ \hline
9 & $\mathrm{N(0,1)}$ & 5\%  & 0.082 (0.015)  & 0.050 (0.015) & 0.055 (0.030) & 0.038 (0.011)  & \textbf{0.020} (0.005)  \rule[0.5ex]{0pt}{2ex}\\
  &    & 10\% & 0.143 (0.018)  & 0.108 (0.016)  & 0.050 (0.025)  & 0.065 (0.011)  & \textbf{0.024} (0.005)  \rule[0.5ex]{0pt}{2ex}\\
  &    & 20\% & 0.345 (0.025) & 0.299 (0.022) & 0.073 (0.038)  & 0.183 (0.030)  & \textbf{0.039} (0.007)  \rule[0.5ex]{0pt}{2ex}\\ \cmidrule{2-8} 
  & $\mathrm{t(3)}$ & 5\%  & 0.136 (0.034)  & 0.098 (0.024)  & 0.097 (0.047) & 0.070 (0.015)  & \textbf{0.022} (0.007) \rule[0.5ex]{0pt}{2ex}\\
  &    & 10\% & 0.190 (0.026)  & 0.153 (0.016)  & 0.101 (0.030)  & 0.100 (0.021)  & \textbf{0.033} (0.007)  \rule[0.5ex]{0pt}{2ex}\\
  &    & 20\% & 0.373 (0.041)  & 0.323 (0.034)  & 0.146 (0.098)  &0.203 (0.038)  & \textbf{0.046} (0.008)   \rule[0.5ex]{0pt}{2ex}\\ \hline
\end{tabular}%
}
\caption{The averages and standard errors (in parentheses) of MSE based on 100 replications for Example 2 are presented. The rate of outlier of $\alpha$ is set to 5\%, 10\%, and 20\%. The smallest MSE is written in bold.}
\label{tbl:ex2}
\end{table}

\subsection{Example 3}\label{sim:ex3}
To check the variable selection performance of the (A)GDNN-LD, the following nonlinear sparse multivariate regression model is considered.
\begin{equation}\label{ex3}
    y_{k} =  0.1k + 0.1x_1^3 + 0.5x_2 ^2 + x_1 + x_2 + x_3 + \epsilon_k,
\end{equation} 
where $k=1,2,3$ and $x_p \sim U(-3,3), p=1,2,\cdots,10$, are independent of each other. For the random error vector $\be = (\epsilon_1,\epsilon_1, \cdots, \epsilon_k)^{\top}$, we also considered two different error distributions; a multivariate normal distribution (MN) denoted as $N_k (\mathbf{0}, \Sigma_{\epsilon})$, and a multivariate t distribution (MT) with 3 degrees of freedom. $\Sigma_{\epsilon}$ is set to the identity matrix. Three response variables $(y_1, y_2, y_3)$ are only dominated by the first three predictors $(x_1, x_2, x_3)$ and the rest are irrelevant variables.

Table \ref{tbl:ex3} summarizes the result. The result of the Oracle case, which fits the model with only the relevant predictors, is provided as a benchmark. Model selection performance was measured by the number of correctly selected predictor variables on average (NC), the number of incorrectly selected predictor variables on average (NIC), and the number of times that the true model was correctly identified (NT). %If $\left\| \widehat{\bw}_{(j)}^{(1)}\right\|^2 \le 0.5\times10^{-3}$ for $j=0,\cdots,p$, then we consider the $j$th  predictor $x_j$ as an irrelevant variable under the model. 
Moreover, in the last column of Table \ref{tbl:ex3} the Frobenius norm, formulated in \eqref{fnorm}, between the true coefficients matrix, which is a zero matrix $\mathbf{B}_0=[\mathbf{0} \vert \mathbf{0} \vert \mathbf{0}] \in \mathbb{R}^{7 \times 3}$ for the irrelevant predictors $(\bx_4, \cdots, \bx_{10} )$ and the estimated coefficient matrix $\widehat{\mathbf{B}}_0=\left[\hat{\boldsymbol{\beta}}_4 \vert \hat{\boldsymbol{\beta}}_5 \vert \cdots \vert \hat{\boldsymbol{\beta}}_{10}\right]^T \in \mathbb{R}^{7 \times 3}$ is presented. 
\begin{equation}\label{fnorm}
\left\|\mathbf{B}_0-\widehat{\mathbf{B}}_0\right\|_F=\sqrt{\sum_{i=1}^m \sum_{j=1}^n\left|b_{i j}-\hat{b}_{i j}\right|^2},
\end{equation}
where $b_{ij}$ means  $(i,j)$th element of the matrix ${\mathbf{B}}_0$. 
As shown in Table \ref{tbl:ex3}, the AGDNN-LD shows the comparable prediction power to the Oracle's result. Also, the AGDNN-LD estimator tends to estimate a sparser model than the GDNN-LD estimator for every simulation setting.

\begin{center}
\begin{table}[t]
\resizebox{\textwidth}{!}{%
\begin{tabular}{cclccccc}
\hline
\multirow{2}{*}{$\be$} & \multirow{2}{*}{$\alpha$} & \multirow{2}{*}{Method} & \multirow{2}{*}{Ave. MSE} & \multicolumn{2}{c}{No. of variables selected} & \multirow{2}{*}{NT} & \multirow{2}{*}{\begin{tabular}[c]{@{}c@{}}Frobenius norm\\ ($\times 10^{-3}$)\end{tabular}} \\ \cmidrule{5-6} 
    &       &                         &                      &  NC                   & NIC                    &                     &            \rule[0ex]{0pt}{2ex}   \\ \hline
MN  & 0\%    & Oracle      &0.119 (0.008)    &3     &  0  & 100   &   0   \rule[0.5ex]{0pt}{2ex}\\
    &      & DNN-LD       & 1.041 (0.025)      & 3    &  7    & 0   &3.443 (0.477)\rule[0.5ex]{0pt}{2ex}\\
    &      & GDNN-LD       & 0.249 (0.008)     &  3   &  3.760 (0.221)   &  5   &0.017 (0.008) \rule[0.5ex]{0pt}{2ex}\\
    &      & AGDNN-LD      & \textbf{0.128} (0.006)     & 3   &  \textbf{0.980} (0.101)   & \textbf{38}   & \textbf{0.001} (0.001) \rule[0.5ex]{0pt}{2ex} \\ \cmidrule{2-8}
    & 10\%   & Oracle      & 0.138 (0.005)  & 3    & 0   & 100   & 0\\
    &      & DNN-LD       & 1.114 (0.015)   & 3 &  7   & 0   &  3.236 (0.500) \rule[0.5ex]{0pt}{2ex}\\
    &      & GDNN-LD       & 0.279 (0.006)    & 3   & 3.460 (0.187)   & 6   & 0.0026 (0.001) \rule[0.5ex]{0pt}{2ex}\\
    &      & AGDNN-LD      & \textbf{0.143} (0.006)      & 3    & \textbf{0.820} (0.093)   & \textbf{46}   & \textbf{0.001} (0.001) \rule[0.5ex]{0pt}{2ex}\\ \hline
MT  & 0\%   & Oracle       &0.113 (0.009)  &3     &0     &100    &0  \rule[0.5ex]{0pt}{2ex}\\
    &     & DNN-LD         & 1.358 (0.023)    &3     & 7    & 0   & 4.887 (0.726)    \rule[0.5ex]{0pt}{2ex}\\
    &     & GDNN-LD        & 0.236 (0.009)   &  3   & 5.124 (0.127)    &  2  &0.004 (0.001) \rule[0.5ex]{0pt}{2ex}\\
    &      & AGDNN-LD      & \textbf{0.123} (0.003)   &  3   & \textbf{1.312} (0.052)    & \textbf{17}  & \textbf{0.001} (0.000) \rule[0.5ex]{0pt}{2ex}\\ \cmidrule{2-8} 
    & 10\%    & Oracle     & 0.183 (0.007)    & 3    & 0    & 100   & 0  \rule[0.5ex]{0pt}{2ex}\\
    &      & DNN-LD        &1.408 (0.028)     & 3     & 7   & 0   & 3.011 (0.356)  \rule[0.5ex]{0pt}{2ex}\\
    &      & GDNN-LD       & 0.321 (0.005)    & 3    & 3.190 (0.182)  & 5   & 0.004 (0.002) \rule[0.5ex]{0pt}{2ex}\\
    &      & AGDNN-LD      & \textbf{0.197} (0.009)    & 3    & \textbf{1.160} (0.114)  & \textbf{33}   & \textbf{0.001} (0.001) \rule[0.5ex]{0pt}{2ex}\\ \hline
\end{tabular}%
}
\caption{The averages and standard errors (in parentheses) of MSE, NC, NIC, NT, and Frobenius norm based on 100 replications are summarized in the table. The result with $\alpha=10\%$ of the outlier is also investigated. The best results except for the case of the oracle are written in bold.}
\label{tbl:ex3}
\end{table}
\end{center} 

\section{Application to real data analysis}\label{sec5}

In this section, we applied the DNN-LD and its penalized methods to the concrete slump test data. Concrete slump test data \citep{Yeh2009data} is publicly available through UCI Machine Learning Repository. The data have 103 observations with Flow $(y_1)$, Slump $(y_2)$, and Compressive strength $(y_3)$ as response variables and Cement $(x_1)$, Slag $(x_2)$, Fly ash $(x_3)$, Water $(x_4)$, Super plasticizer $(x_5)$, and Coarse aggregate $(x_6)$ as 6 signal predictors. Distance correlation among response variables is $\overline{dcor}(y_1, y_2)=0.84$, $\overline{dcor}(y_1, y_3)=0.34$ and $\overline{dcor}(y_2, y_3)=0.34$ on average.

To confirm the model selection performance of the proposed estimator, we intentionally add two noise variables,  $x_8$ and $x_9$, which are generated randomly from $U(-1,1)$ distribution, to the nonlinear regression model. We split the data into training and test data with the proportion of 70\%, and 30\%, respectively. The result of 100 replication from 5-fold cross-validation is summarized in Table \ref{tbl:real_data} with the metrics of Mean Squared Prediction Error (MSPE), averages of number of selected signal and noise variables, and Frobenius norm in $10^{-3}$ scale. 10\% of outlier setting is also considered by adding 5 to each response variable.
The result echoes the outcome of the simulations in the previous Section \ref{sec4}. %The AGDNN-LD shows its superiority over the other methods both in terms of model prediction accuracy and model selection criteria for two different levels of outlier settings. 

\begin{center}
\begin{table}
\resizebox{\textwidth}{!}{%
\begin{tabular}{clcccc}
\hline
\multirow{2}{*}{$\alpha$} & \multirow{2}{*}{Method} & \multirow{2}{*}{MSPE} & \multicolumn{2}{c}{No. of variables selected} & \multirow{2}{*}{Frobenius norm } \\ \cmidrule{4-5}
                   &                         &                       & signal                   & noise                   & ($\times 10^{-3}$)    \rule[0ex]{0pt}{2ex} \\ \hline
0\%             & DNN-LD      &  0.489 (0.174) &  6    & 2    &   0.040 (0.005)  \rule[0.5ex]{0pt}{2ex}  \\
                   & GDNN-LD      & 0.419 (0.139)     &  5.980 (0.141)   &  1.910 (0.362)   &  0.003 (0.001)  \rule[0.5ex]{0pt}{2ex}\\
                   & AGDNN-LD     & 0.387 (0.105)   & 5.970 (0.171)   &  1.760 (0.474)   & 0.002 (0.001)    \rule[0.5ex]{0pt}{2ex}\\ \hline                   
10\%           & DNN-LD        & 0.559 (0.194)   & 6   & 2     & 0.053 (0.008)         \rule[0.5ex]{0pt}{2ex}\\
                   & GDNN-LD       & 0.446 (0.125)   & 5.970 (0.171)   &  1.870 (0.418)    & 0.002 (0.001)    \rule[0.5ex]{0pt}{2ex}\\
                   & AGDNN-LD     &  0.431 (0.124)   & 5.910 (0.288)   & 1.670 (0.553)  & 0.002 (0.001)        \rule[0.5ex]{0pt}{2ex}\\ \hline
                   
\end{tabular}%
}
\caption{The Average and standard deviations of MSPE, number of variables selected, and Frobenius norm based on 100 replications with 5-fold cross-validation are summarized.}
\label{tbl:real_data}
\end{table}
\end{center}

\section{Concluding remarks}\label{sec6}
In this study, we propose a deep neural network (DNN) based least distance (LD) estimator (DNN-LD) for a multivariate regression problem. In order to exploit the sharing strength among correlated responses in the estimation, we used the least distance loss and applied a DNN architecture to enjoy its modeling flexibility. The numerical studies showed that the prediction performance of DNN-LD is promising when the correlation among response variables increases and the situation of the presence of the outliers.
Moreover, by introducing a group Lasso penalty to the first weight matrix of the neural network, we suggested GDNN-LD and AGDNN-LD for achieving a variable selection property. In addition, in terms of computation, we applied quadratic smoothing approximation method to overcome the non-differentiability problem of the least distance loss function. A real data analysis of concrete slump test data confirm the usefulness of the proposed methods.

By the way, In this paper, we only assert our proposed methods via a couple of simulations and real data analysis. It is worthwhile to analyze the theoretical properties of the proposed methods, which have been partially studied by \cite{Wang2016} and \cite{Zhang2019}. These parts are left as further research topics.

\vspace{.5cm}

\textbf{Acknowledgements} S. Bang and J. Shin's work was supported by Basic Science Research Program through the National Research Foundation of Korea (NRF) funded by the Ministry of Science and ICT (Grant No. NRF-2022R1F1A1061622).

\bibliography{references}

\end{document}